\documentclass[twocolumn,showpacs,preprintnumbers,amsmath,amssymb,superscriptaddress]{revtex4}
\usepackage{graphicx}
\usepackage{dcolumn}
\usepackage{bm}
\usepackage{soul}
\usepackage{color}
\usepackage{epstopdf}
\usepackage[version=3]{mhchem}
\usepackage{lipsum}
\usepackage[outercaption]{sidecap}
\usepackage{floatrow}
\begin{document}

\title{Electric potential profile of a spherical soft particle with a charged core}
\author{Anh D. Phan}
\affiliation{Institute of Physics, 10 Daotan, Badinh, Hanoi, Vietnam}%
\email{anhphan@mail.usf.edu}
\author{Dustin A. Tracy}
\affiliation{Department of Physics, University of Florida, Gainesville 32611, USA}
\author{T. L. Hoai Nguyen}
\affiliation{Institute of Physics, 10 Daotan, Badinh, Hanoi, Vietnam}
\author{N. A. Viet}
\affiliation{Institute of Physics, 10 Daotan, Badinh, Hanoi, Vietnam}

\author{The-Long Phan}
\affiliation{Department of Physics, Chungbuk National University, Cheongju 361-763, Korea}
\email{ptlong2512@yahoo.com}
\author{Thanh H. Nguyen}
\affiliation{Department of Civil and Environmental Engineering, University of Illinois at Urbana-Champaign, Urbana-Champaign, USA}
\date{\today}

\begin{abstract}
The electrostatic potential profile of a spherical soft particle is derived by solving the Poisson-Boltzmann equations on a spherical system both numerically and analytically. The soft particle is assumed to consist of an ion-permeable charged outer layer and a non-permeable charged core with constant charged density. The contribution of the core to the potential profile is calculated for different charges and dielectric constants. Our results show that the charged core heavily influences the local potential within the soft particle. In contrast, the potential distribution outside the particle in the salt solution is found to be weakly dependent on the core features. These findings are consistent with previous experiments showing the minor impact of the core of the MS2 virus on its overall electrical properties. Our studies also indicate that while a change in temperature from 290 $K$ to 310 $K$ only slightly varies the potential, the ionic strength in the range of 1-600 mM has a significant effect on the potential profile. Our studies would provide good understanding for experimental research in the field of biophysics and nanomedicine.
\end{abstract}

\maketitle
\section{Introduction}
In recent years, the rapid advancement of nanotechnology has recently opened up novel proposals for the field of soft particles due to their wide range of applications in life science and material science \cite{1,2,3} causing the study of soft particles to be of central interest to interdisciplinary areas combining physics, chemistry and biology. Despite much effort to theoretically understand the properties of soft particles \cite{1,4,5,6,7}, the theoretical models still face a variety of problematic issues and challenges coming from the complexity of biological structures and the variation of solvents. The theoretical approaches allow us to interpret the particle systems in a simplified manner. Thus, the construction and development of models to help explain new phenomena and experiments are important to the further understanding of these systems.

Introduced for the first time in 1994 \cite{7} and then advanced mostly by the works of H. Ohshima \cite{4,5,6,8}, the theory of soft particle (SPE) provides a powerful tool for investigating the behavior of biocolloidal particles, including  bacteria and viruses. The main results produced by the theory are the electrostatic and electrokinetic properties of soft particles with different geometrical shapes immersed in electrolyte solutions. In this model, the soft particles are described to have a non-penetrable neutral hard core coated by an ion permeable polyelectrolyte soft layer with negative constant volume density charge \cite{5,6,8}. The electric potential distribution is obtained by solving the Poisson-Boltzmann equations. The electrophoretic mobility is a solution of the Navier-Stokes equation for the velocity of the liquid flow through the permeable layer. Moreover, the model can be exploited to study the interactions between soft particles. The prediction of non-zero mobility for soft particles in solutions with low ionic strength obtained by the Ohshima model was verified by experimental data \cite{9,10}. The development for the Ohshima model has been done by taking into account the effect of inhomogeneous charge distribution in the soft layer. In the diffuse soft particle electrokinetic theory (DSPE) given by J. F. Duval \cite{11,12}, the gradual decay of the electrodynamic properties of the soft layer inside and outside the particle was considered. The extensions of the DSPE theory to multilayer soft particles have been recently done \cite{13,14}. The inhomogeneous distribution of the charged polymer segments is demonstrated to substantially affect the overall electrokinetic response, especially in the low electrolyte concentration regime. 

While the surface charge density has been deeply investigated, the core charge distribution has been rarely taken into account. In most cases, a core charge is assumed to be zero, so the electrical potential inside the core remains unchanged. A theoretical study \cite{20,19,21} mentioned the charge of the virus core in general cases to calculate the nonspeciﬁc electrostatic interactions in virus systems. The ratio between the volume charge density of the core and that of the surface layer is measured to be half of that found in bacteriophage MS2 \cite{15} suggesting that the effect of the core charge on the electrostatic and electrokinetic properties of the particle should be re-examined. %However, small-angle neutron scattering on microgel particles demonstrated that there was non-zero charge distribution from the center of the particle to its periphery []. For example, the ratio between volume charge density of core and that of the surface layer was mesured to be a half experimentally with bacteriophage MS2 [10] so that the effect of the core charge on the electrostatic and electrokinetic properties of the particle should be re-examined. 

In the present work, we propose a new simple core-shell model for a soft particle based on the assumption that the soft particle now consists of a charged hard core with a volume charge density and a charged outer layer. We study the contribution of the core parameters, such as the core charge and the core dielectric constant, to the particle's electrostatic properties. Our calculations provide the first theoretical investigation about the effects of temperature and salt concentration on the electrostatic properties.
\section{Theoretical background}

\begin{figure}[htp]
\includegraphics[width=8cm]{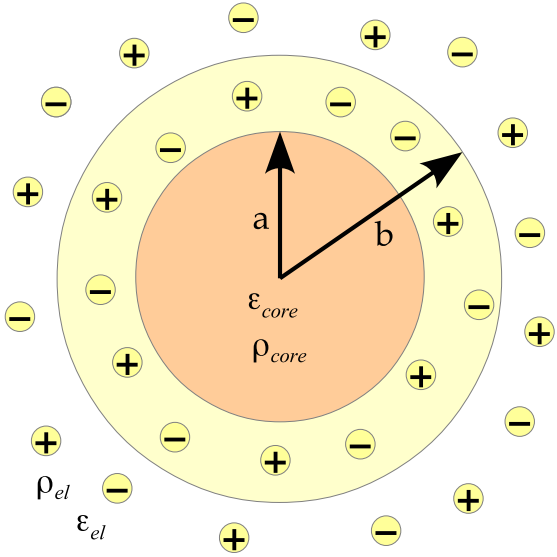}% Here is how to import EPS art
\caption{\label{fig:1}(Color online) The theoretical model of a soft particle including a hard core with the charge density $\rho_{core}$ and the dielectric constant $\varepsilon_{core}$, and an ion-penetrable surface layer of polyelectrolytes coated around. The soft particle is immersed in an electrolyte solution with the charge density $\rho_{el}$ and the permittivity $\varepsilon_r$.}
\end{figure}

We consider a soft particle with radius $b$ immersed in an electrolyte solution. The soft particle is assumed to contain a hard core of radius $a$ coated by an ion-penetrable surface charge layer of polyelectrolyte with thickness $(b-a)$. Identified with the Ohshima model \cite{5,6,8}, the volume charge density of the soft shell is $ZNe$, where $e$ is an electron charge, $Z$ and $N$ are the valence and the charge density of the polyelectrolyte ions, respectively. In our work, the neutral hard core in the Ohshima model is substituted by the charged core with a constant volume charge density $\rho_{core}$ and a dielectric constant $\varepsilon_{core}$ (see in Fig. \ref{fig:1}).

The electric potential distribution obeys the Poisson–Boltzmann equations

\begin{eqnarray}
\left\{ \begin{array}{rcl}
\triangle\psi = -\cfrac{\rho_{el}}{\varepsilon_r\varepsilon_0}, & \
& b \le r < \infty \\
\triangle\psi = -\cfrac{\rho_{el}+ZNe}{\varepsilon_r\varepsilon_0}, &  & a \le r < b 
\\ \triangle\psi = -\cfrac{\rho_{core}}{\varepsilon_{core}\varepsilon_0}, &  & 0 \le r < a
\end{array}\right.
\label{1}
\end{eqnarray}
where $\varepsilon_r$ is the permittivity of the aqueous solution, $\rho_{el}$ is the charge density of the ions in solvent and can be described by the Boltzmann distribution \cite{5,8}
\begin{eqnarray}
\rho_{el}(r) = \sum_{i=1}^M z_i e n_i exp\left(-\cfrac{z_i e \psi}{k_BT}  \right),
\label{2}
\end{eqnarray}
$M$, $z_i$, $n_i$ are the number ion types, the $i$\emph{th} ionic valance and the ion concentration in solution, respectively. In our calculations, we consider a simple case where an aqueous solution only contains a monovalent salt $M = 2$ and $z_i=\left\lbrace -z,z \right\rbrace$, and therefore
\begin{eqnarray}
\rho_{el}(r) = -2nze\sinh\left(\cfrac{ze\psi}{k_BT}\right).
\label{3}
\end{eqnarray}

The spherical Poisson-Boltzmann equation in Eq.(\ref{1}) does not have a general analytical solution. With rapid development of numerical methods, algorithms and tools, Eq.(\ref{1}) can be numerically solved. Using several approximations, however, provide analytical expressions that well explain physical phenomena. In the case of a low potential, the charge density in the electrolyte solution is given by
\begin{eqnarray}
\rho_{el}(r) = -\frac{2nz^2e^2}{k_BT}\psi.
\label{4}
\end{eqnarray}

Substituting Eq.(\ref{4}) into Eq.(\ref{1}) provides
\begin{eqnarray}
\left\{ \begin{array}{rcl}
\cfrac{d^2\psi}{dr^2}+\cfrac{2}{r}\cfrac{d\psi}{dr} = \kappa^2\psi, & \
& b \le r < \infty \\
\cfrac{d^2\psi}{dr^2}+\cfrac{2}{r}\cfrac{d\psi}{dr} = \kappa^2\left(\psi -\cfrac{ZNe}{\kappa^2\varepsilon_r\varepsilon_0}  \right), &  & a \le r < b 
\\ \cfrac{d^2\psi}{dr^2}+\cfrac{2}{r}\cfrac{d\psi}{dr} = -\cfrac{\rho_{core}}{\varepsilon_{core}\varepsilon_0}, &  & 0 \le r < a
\end{array}\right.
\label{5}
\end{eqnarray}
where $\kappa^2 = 2z^2e^2n/\varepsilon_r\varepsilon_0k_BT$ is the Debye-Huckel parameter \cite{8}. The general solution of Eq.(\ref{5}) give us% The potential in our system obey the following boundary conditions
\begin{eqnarray}
\left\{ \begin{array}{rcl}
\psi(r) = A_1\cfrac{e^{-\kappa r}}{r} + B_1\cfrac{e^{\kappa r}}{r}, & \
& b \le r < \infty \\
\psi(r) = A_2\cfrac{e^{-\kappa r}}{r} + B_2\cfrac{e^{\kappa r}}{r} + \cfrac{ZNe}{\kappa^2\varepsilon_r\varepsilon_0} , &  & a \le r < b 
\\ \psi(r) = \cfrac{A_3}{r}+B_3-\cfrac{1}{6}\cfrac{\rho_{core}r^2}{\varepsilon_{core}\varepsilon_0}. &  & 0 \le r < a
\end{array}\right.
\label{6}
\end{eqnarray}

The coefficients $A_1$, $A_2$, $A_3$, $B_1$, $B_2$ and $B_3$ in Eq.(\ref{6}) can be found by applying the following  boundary conditions:
\begin{eqnarray}
\psi(\infty) = 0&,& \psi(0) \neq \infty, \label{7} \\
\psi(a^-) = \psi(a^+)&,& \psi(b^-) = \psi(b^+), \label{8} \\
\varepsilon_{core}\varepsilon_0\psi '(a^-) = \varepsilon_{r}\varepsilon_0\psi '(a^+)&,& \psi '(b^-) = \psi '(b^+). \label{9}
\end{eqnarray}

In the previous studies \cite{5,6,8}, the condition $\psi '(a^-) = 0$ is used due to the assumption that the electric fields inside the core must be equal to zero and so the electric potential is set to be constant. However, this assumption is valid only if the core is composed of a metal material. In the general case, the continuity of electric displacement fields has to be applied.

%\end{itemize}
\section{Numerical results and discussions}
The analytical expressions of $\psi(r)$ can be derived after some straightforward calculations. To understand the potential contribution of the charged hard core $\psi_{core}(r)$ and the soft shell $\psi_{shell}(r)$ to the total potential, $\psi(r)$ can be writen as the sum of two terms $\psi(r)=\psi_{core}(r)+\psi_{shell}(r)$
\begin{eqnarray}
\psi_{core}= \left\{ \begin{array}{rcl}
\cfrac{1}{3\varepsilon_r\varepsilon_0}\cfrac{\rho_{core}a^3}{r}\cfrac{e^{-\kappa(r-a)}}{1+\kappa a}, &  a \le r < \infty \\ 
\cfrac{\rho_{core}(a^2-r^2)}{6\varepsilon_{core}\varepsilon_0}+\cfrac{\rho_{core}a^2}{3\varepsilon_{r}\varepsilon_0(1+\kappa a)}. & 0 \le r < a
\end{array}\right.
\label{10}
\end{eqnarray}

\begin{widetext}
\begin{eqnarray}
\psi_{shell}= \left\{ \begin{array}{rcl}
\cfrac{ZNe}{2\varepsilon_r\varepsilon_0\kappa^2}\left[1-\cfrac{1}{\kappa b}+\cfrac{(1-\kappa a)(1+\kappa b)}{1+\kappa a}e^{-2\kappa (b-a)} \right]\cfrac{be^{-\kappa (r-b)}}{r}, &  b \le r < \infty \\ 
\cfrac{ZNe}{\varepsilon_r\varepsilon_0\kappa^2}\left[1-\cfrac{1+\kappa b}{1+\kappa a}e^{-\kappa (b-a)}\left(\cfrac{\sinh \kappa(r-a)}{\kappa r}+ \cfrac{a\cosh \kappa(r-a)}{r} \right)\right], & a \le r < b \\
\cfrac{ZNe}{\varepsilon_r\varepsilon_0\kappa^2}\left[1-\cfrac{1+\kappa b}{1+\kappa a}e^{-\kappa (b-a)}\right]. & 0 \le r < a
\end{array}\right.
\label{11}
\end{eqnarray}
\end{widetext}

\begin{figure}[htp]
\includegraphics[width=8.8cm]{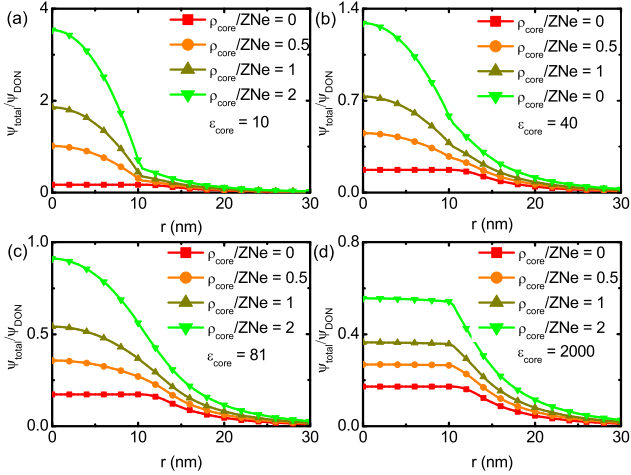}% Here is how to import EPS art
\caption{\label{fig:2}(Color online) The potential of the bacteriophage MS2 immersed in NaCl solution as a function of distance $r$ for different values of $\rho_{core}/ZNe$ with a) $\varepsilon_{core} = 10$, b) $\varepsilon_{core} = 40$, c) $\varepsilon_{core} = 81$, and d) $\varepsilon_{core} = 2000$ at 298 $K$.}
\end{figure}

Note that when $\rho_{core} = 0$, $\psi_{core}(r)=0$ and $\psi(r)=\psi_{shell}(r)$ for $r \geq a$, the expressions are exactly identical to the potential obtained by Ohshima in Ref.\cite{5,6,8}. This finding shows that our calculations are suitable with results found in particular cases. The potential outside the soft particle $(r \geq b)$ can be expressed via the surface potential, which is defined as $\psi_S = \psi(b)$,
\begin{eqnarray}
\psi (r) = \psi_S\frac{b}{r}e^{-\kappa(r-b)}.
\label{12}
\end{eqnarray}

Equation (\ref{12}) suggests that the potential profile can be obtained from understanding of the surface potential.

Figure \ref{fig:2} shows the normalized electrostatic potentials of the bacteriophage MS2 as a function of distance from the center of the soft particle with various values of the core charge. The selected ratios $\rho_{core}/ZNe$ in our calculations are positive because previous studies \cite{11,12,15} claimed that the charges of both core and shell have the same sign. $\psi_{DON}=ZNe/\varepsilon_r\varepsilon_0\kappa^2$ is the linearized Donnan potential \cite{5}. The salt solution around MS2 is 1 mM NaCl. The values of radius $a = 10.3$ nm and $b = 13.6$ nm were chosen from the experiment \cite{15}. The temperature- and concentration-dependent permittivity of the solvent is given by \cite{16}
\begin{eqnarray}
\varepsilon_r &\equiv& \varepsilon_r(C,T) = \varepsilon_W(T)h(C),\nonumber\\
\varepsilon_W(T) &=& 249.4 - 0.788T + 7.2\times 10^{-4}T^2,\\
h(C) &=& 1 - 0.255C + 5.15\times 10^{-2}C^2-6.89\times 10^{-3}C^3,\nonumber
\label{13}
\end{eqnarray}
where $C$ is the ionic strength of the solution.

It is important to note that the experiments in Ref.\cite{15} are carried out at pH = 5.9 while the expression of $\varepsilon_r(C,T)$ is valid for the NaCl solution at a pH of 7. The measurements in the previous work \cite{15}, however, reveals that the electrophoretic mobility remains nearly unchanged in the pH range of 6 to 8. Another previous study \cite{18} demonstrates that the thickness of a protein layer coated on surface of a gold nanoparticle is approximately constant at the same frequency range. The size of radius $a$ and $b$ , therefore, can be assumed to not be influenced by the pH in this regime, and Eq.(\ref{13}) is well-described for the MS2 solution in Ref.\cite{15}.

\begin{figure*}[htp]
\includegraphics[width=16.5cm]{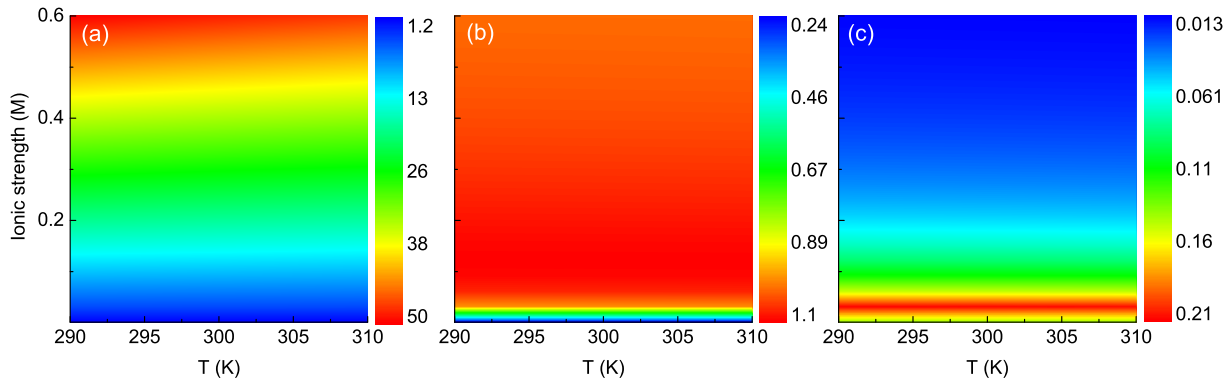}
\caption{\label{fig:3}(Color online) Ionic strength vs. temperature contour plots of the normalized electrostatic potentials $\psi_{total}/\psi_{DON}$ for the bacteriophage MS2 with $\varepsilon_{core} = 40$ and $\rho_{core}/ZNe = 0.5$ at a) $r = 8$ nm, b) $r = 12$ nm, c) $r = 15$ nm.}
\end{figure*}

The potential profiles with $\varepsilon_{core}$ different from the value for the surrounding medium are not smooth and contain a breaking point at the boundary between the core and the outer layer. When the core dielectric constant is close to the permittivity of the solvent ($\varepsilon_r \approx 79$), the curves of potential become smooth as shown in Fig. \ref{fig:2}c. The soft shell may be thin enough for the potential to not change suddenly at the interface of the particle ($r = b$). The increase of the core charge leads to the rise of the potential profile of the soft particle, particularly area inside the particle. This observation indicates the large effect of the charge and the permittivity of the hard core on $\psi(r)$. Outside the hard core, the potential induced by the core charge exponentially decays and is negligible at long distances. This finding can be explained via Eq.(\ref{11}), $\psi_{core}$ is proportional to $e^{-\kappa(r-a)}$ when $(r \geq a)$. This result also helps us to understand the similarity of MS2 with and without the RNA core in experiment \cite{15}. Increasing $\varepsilon_{core}$ causes a reduction of the ratio $\psi_{total}/\psi_{DON}$. If the core dielectric constant is sufficiently large ($\varepsilon_{core}=2000$), the core potential is independent of distance $r$ due to the cancellation of the term $\rho_{core}(a^2-r^2)/6\varepsilon_{core}\varepsilon_0$.

The impact of the NaCl concentration and temperature on the potential of virus nanoparticles MS2 are shown in Fig. \ref{fig:3}. It is well-known that the thermal effect causes DNA and protein denaturation \cite{17,18}. At around 350 - 360 $K$, these biological systems transition between conformations and links between strands broken \cite{17,18}. The variation of a soft particle's size as a function of temperature is an interesting topic but one that is still poorly understood. Authors in Ref.\cite{17} indicate that the localized surface plasmon resonance of protein-coated gold nanoparticles with diameter 13 nm is nearly unchanged at temperatures ranging from 290 to 310 $K$. As a result, we suppose that the sizes of the core and the soft layer does not vary much in the temperature regime used in our calculations. The same assumption is provided with the effect of ionic strength on the size of the MS2 virus.

Although a temperature increase creates more space between the molecules of the solutions and is a reason for a decrease in $\varepsilon_r$ at certain ionic strength, the presence of $\sqrt{T}$ in the denominator of the expression for $\kappa$ causes the enhancement of $\kappa$ and the slight reduction in the potential inside the core (see Fig.\ref{fig:3}a) via the exponential function. However, the thermal effects on $\psi_{total}$ in the soft shell and outside the particle has little effect on the system. The result suggests that the thermal factor cannot be used to tailor the total potential. 

Interestingly, the potential is quite sensitive to the change of ionic strength at fixed temperatures. Ions in the solution alter the dielectric response of the solvent in comparison with pure water. $\psi_{total}$ in the core and shell of a particle dramatically increases when the salt concentration is varied from 1 mM to 600 mM. At larger ionic strength, $\kappa$ increases and the exponential functions in Eq.(\ref{10}) and (\ref{11}) are quickly cancelled out in the area $ r \le b$. As can be seen in Fig. \ref{fig:3}b, $\psi_{total}/\psi_{DON} \approx 1$ at $C \geq 40$ mM. Meanwhile, the normalized potential at $r = 8$ nm extends to  50 when $C = 0.6$ M since $\psi_{total}/\psi_{DON} \sim \varepsilon_r\kappa^2 \sim n$. For this reason, the results coming from Fig.(\ref{fig:3})a and b can be easily understood. For $r \geq b$, $\psi_{total}$ is mainly contributed by $\psi_{shell}$. Equation \ref{11} expresses that the shell potential is proportional to $e^{-k(r-b)}$. We, therefore, can observe the diminishment of the normalized potential as the ionic strength increases in Fig. \ref{fig:3}c.

\section{Conclusions}
In conclusion, the potential profile of a soft particle with a charged hard core has been investigated in detail with various values of the charge and dielectric constant. Our results show that the potential inside the core is strongly modified in comparison with the case of the neutral core introduced by the Ohshima model. However the potential induced by the core charge nearly disappears outside the particle and the total potential is determined by the electric properties of the soft shell. Our model succeeds in explaining the identical properties of untreated- MS2 and RNA-free MS2 reported in previous works. We also take into account the effects of temperature and ionic strength on the potential. It is demonstrated that while the thermal effect can be ignored in the temperature regime of 290-310 $K$, the salt concentration still has significant influence.  %This finding proves the weak influence of the core on the total potential outside particle. 
\begin{acknowledgments}
This work was supported by the Nafosted Grant No. 103.01-2013.25. THN is partially supported by USDA NIFA grant 2013-67017-21221.
\end{acknowledgments}

\end{document}